# Optimization of the coherence properties of diamond samples with an intermediate concentration of NV centers

Rubinas O.R.[1,2,3], Soshenko V.V.[2,3], Bolshedvorskii S.V.[1,2,3], Zeleneev A.I.[1], Galkin A.S. [4,5], Tarelkin S.A.[4,5], Troschiev S.Y. [4,5], Vorobyov V.V [6], Sorokin V.N.[2], Sukhanov A.A.[7], Vins V.G[8], Smolyaninov A.N.[3] Akimov A.V.[2,9]

[1]Moscow Institute of Physics and Technology, Dolgoprudniy, Moscow Region, Russia

[2]P. N. Lebedev Physical Institute of the Russian Academy of Sciences, Moscow, Russia

[3]LLS Sensor Spin Technologies, Moscow, Russia

[4]Technological Institute for Superhard and Novel Carbon Materials, Troitsk, Moscow 108840, Russia

[5]The All-Russian Research Institute for Optical and Physical Measurements, Moscow 119361, Russia

[6]The University of Stuttgart, Stuttgart 70049, Germany

[7]Zavoisky Physical-Technical Institute, FRC Kazan Scientific Center of RAS, Kazan 420029, Russia

[8]LLS Velman, Novosibirsk, Russia

[9]Texas A&M University, 4242 TAMU, College Station, USA

## Corresponding author:

Alexey V Akimov, akimov@physics.tamu.edu

## Abstract:

The sensitivity of the nitrogen-vacancy (NV) color centers in diamond-based magnetometers strongly depends on the number of NV centers involved in the measurement. Unfortunately, an increasing concentration of NV centers leads to decreases of their dephasing and coherence time if the nitrogen content exceeds a certain threshold level (approximately $10^{17}$ cm$^{-3}$). Here, we demonstrate that this increased dephasing can be efficiently compensated by postprocessing procedures in the vicinity of the threshold concertation, thus extending possible useful concentrations on NV centers with the maximum possible decoherence time in diamonds with a natural carbon content.

## Introduction

One application of nitrogen-vacancy(NV) centers in diamonds is to achieve either a very high resolution or very low magnetic field magnetometry[1]. The ultimate sensitivity of the sensor crucially depends on the amount of collected signal. To this end, NV center ensembles are often advantageous over single color centers[2]. Increasing the density of NV centers is nevertheless impossible without increasing the decoherence rate, which includes increasing the nitrogen-related impurity concentration and some level of interaction of the NV centers themselves[3].

In samples with sufficient purity, there only 2 major sources of decoherence (see Figure 2A,B). The $^{13}C$ – carbon isotope has nuclear spin ($I = 1/2$), is naturally present in diamonds (approximately 1%) and is unavoidable for NV center formation nitrogen, which forms a number of color centers[4]. Among those color centers, a spin containing one is a nitrogen donor (C-defects) with electron spin ($S = 1/2$)[5]. The contribution of these two main sources of decoherence strongly depends on the nitrogen concentration in the sample used. $^{13}C$ defects are more important in diamonds with a low substitutional nitrogen donor concentration of less than $10^{17}$ cm$^{-3}$ (0.6 ppm). At this concentration level, $^{13}C$ becomes the major impurity unless an isotopically pure diamond is used[3]. Here, the dephasing time appears to be independent of the nitrogen content and is basically the highest possible constant for non-isotopically pure diamonds. However, since the substitutional donor nitrogen and NV center concentrations are closely connected, such a regime does not allow for high NV center concentrations[6]. Applications in magnetometry and other sensing application signals within a fixed sample volume requires an increase of the overall nitrogen concentration, which unavoidably leads to an increase of the substitutional nitrogen donor concentration. At the high concentration limit (over $10^{18}$ cm$^{-3}$), substitutional nitrogen donors becomes the main source of decoherence[3], which to a high degree, assuming a constant conversion nitrogen efficiency to the NV center, fully compensates for concentrations related with magnetometry sensitivity. Nevertheless, the higher overall nitrogen concentration, on average, increases the formation efficiency of NV centers, leading to slightly better results at higher concentration[6]. Such an optimum is obviously application dependent[7].

Nitrogen concentration is not the only parameter that affects the conversion efficiency of nitrogen to NV centers. The relative concentrations of $NV^-$, $NV^0$, substitutional donor nitrogen and various other nitrogen related defects in diamonds strongly depend on diamond postprocessing after the diamond has been grown[8]. In particular, if irradiation with an electron beam is used to create vacancies in diamonds and thus increase the $NV^-$ concentration, the electron dose strongly affects the concentration of all the color centers of interest. One of the factors that, to some degree, affects the conversion efficiency is the electron irradiation dose, which is an post-processing parameter[8]. The electron irradiation dose defines the number of vacancies, which during the annealing process, become mobile and thus can convert nitrogen from the substitutional donor into NV centers. Thus, this paper[9] addresses the role of electron irradiation doses over a wide range ($1-100 \cdot 10^{18}$ cm$^{-2}$) for 65 ppm concentrations of initial nitrogen. Somewhat surprisingly, the electron irradiation dose was also found to have a significant effect on the coherence time $T_2$ [10]. In the dose range of $10^{17} - 10^{18}$ cm$^{-2}$, previous researchers found that coherence time could be improved by three times by increasing the irradiation dose. Even the electron energy during electron irradiation has some effect[11].

While much work has been performed already and many aspect of diamond preparation have been carefully studied[3,5–11], the range of the intermediate concentrations of substitutional nitrogen donors, $10^{17} - 10^{18}$ cm$^{-3}$, has not been investigated in detail. This range may be of significant interest since in this regime, if one can reach reasonable conversion efficiencies, the coherence property decrease due to the presence of nitrogen may not be significant, while the NV center concentration may be significant.

In this paper we investigated this unexplored range of diamonds with a borderline nitrogen concentration from $10^{17}-10^{18}$ cm$^{-3}$ (0.6 – 5.6 ppm), where decoherence from substitutional nitrogen donors with $S=1/2$ starts to significantly contribute in the total decoherence rate of NV centers. We explored different postprocessing conditions of diamond plates grown using the low strain high-pressure high-temperature (HPHT) method and demonstrated the possibility of suppressing nitrogen-related decoherence via the efficient conversion of substitutional nitrogen donors into $NV^0$ centers.

## Methods

### Donor nitrogen centers measurements

Diamond plates were gown using the HPHT technique[8]. The samples were irradiated with different electron doses (from $2 \cdot 10^{17}$ through $20 \cdot 10^{17}$ cm$^{-2}$) and annealed at 800°C.

To measure the substitutional nitrogen donors concentration, electron paramagnetic resonance (EPR) spectroscopy[12,13] was used. This method was chosen because, in comparison with infrared (IR) spectroscopy, EPR can be used at concentrations of less than $9 \cdot 10^{17}$ cm$^{-3}$ (5 ppm). The EPR method of concentration metrology[12] requires a normalization sample with a well-known dipole content. To determine the number of donor nitrogen atoms, we compared an EPR signal (Figure 1C) with a diamond sample containing an already known spin content determined via IR spectroscopy[8] (Figure 1A). The concentration of defects was found in the following way: first, the amplitude of the IR resonance frequency absorption of the defect was calculated with respect to the background, and then, the concentration was found with the following formula[8]:

$$n_C [ppm] = \mu_{1130 \text{ cm}^{-1}} \cdot 25 \tag{1}$$

To transform the result into the number of dipoles, we used the following equation:

$$N = n_C [\text{ppm}] \cdot 1.76 \cdot 10^{17} \left[\frac{1}{\text{cm}^3 \text{ppm}}\right] \cdot V[\text{cm}^3] \tag{2}$$

where the coefficient $1.76 \cdot 10^{17}$ cm$^{-3}$/ppm was calculated from the parameters (in volume $V = 1$ cm$^3$, there are $N_A \rho V / m_C$ carbon atoms, the density is $\rho = 3.51 g/cm^3$, the molar mass of carbon is $m_C = 12 g/mol$ and $N_A$ is Avogadro's constant).

For the EPR measurement, the continuous wave (CW)-EPR signal in the first derivative mode was measured. The number of defects of interest could be found from the following double integral[13]:

$$DI = \int_{b1}^{b2} dB \int_{b1}^{B} A(B') dB' = c[G_R C_t n] \frac{\sqrt{P} B_m Q n_B S(S+1) n_S}{f(B_1, B_m)}, \tag{3}$$

where $b_1$ and $b_2$ are magnetic field limits containing the signal of interest, $G_R$ is the receiver gain, $C_t$ is the conversion time, n is the number of scans, $P$ is the microwave power, $B_1$ is the microwave field, $B_m$ is the modulation field, $Q$ is the quality factor of resonator, $n_B$ is the Boltzmann factor, $S$ is the electron spin and $n_S$ is the number of electron spins, and $f$ is a function of the field distribution in the resonator. The number of defects can then be found as follows:

$$N_i = \frac{N_{calibration}}{DI_{calibration}} DI_i \qquad (4)$$

where $DI_{calibration}$ and $N_{calibration}$ are the double integral and the number of defects (measured with IR spectroscopy) for the calibration sample, respectively, and $DI_i$ and $N_i$ are the double integral and the number of defects for sample of interest, respectively.

The number of defects was transformed into density using the diamond plate volume calculated from the mass and diamond density. A low C-defect concentration causes the EPR signal to have high noise, and leads to errors of about 10 to 20% (see Table 1).

To calculate errors in this method, we took into consideration errors in mass determination – $\delta m$ as the average weight error $10^{-4}$ g and IR and EPR spectrum errors as the standard deviation. Thus, there are three sources of errors in the determination of the donor nitrogen concentration based on the following equation:

$$\delta f = \sqrt{\sum \left(\frac{\partial f}{\partial x_i} \delta x_i\right)^2} \qquad (5)$$

For the complex function error, $x_i$ was used for the following three values with errors: mass, the double integral of the EPR spectrum, and the absorption coefficient in the IR spectrum.

### NV centers measurements

However, the EPR method has insufficient accuracy for detecting NV- centers in our diamond sample. Instead, the optical transmission method at a cryogenic temperature of – 77 K was used (Figure 1B). The measurements were carried out on a Vertex 80v Fourier-transform spectrometer (Bruker Optik GmbH, Ettlingen, Germany) with a silicon diode detector. Samples were glued to the substrate with a hole of 0.5 mm and were cooled in the cryostat to the temperature of liquid nitrogen. The concentration of vacancy defects was determined via integration of the corresponding absorption peaks at 637 nm for $NV^-$ and 575 nm for $NV^0$ (Figure 1D)[14] using following equations[15]:

$$\begin{aligned} n[\text{ppb}] &= 5.96 \cdot 10^{-15} \frac{A}{f_{NV}} \\ A &= \frac{1}{d} \int \ln \frac{I_0}{I_1} dE \left[\frac{\text{meV}}{\text{cm}}\right] \\ f_{NV} &= 1.13 \cdot 10^{-16} \left[\text{meV}^2\text{cm}^2\right] \end{aligned} \qquad (6)$$

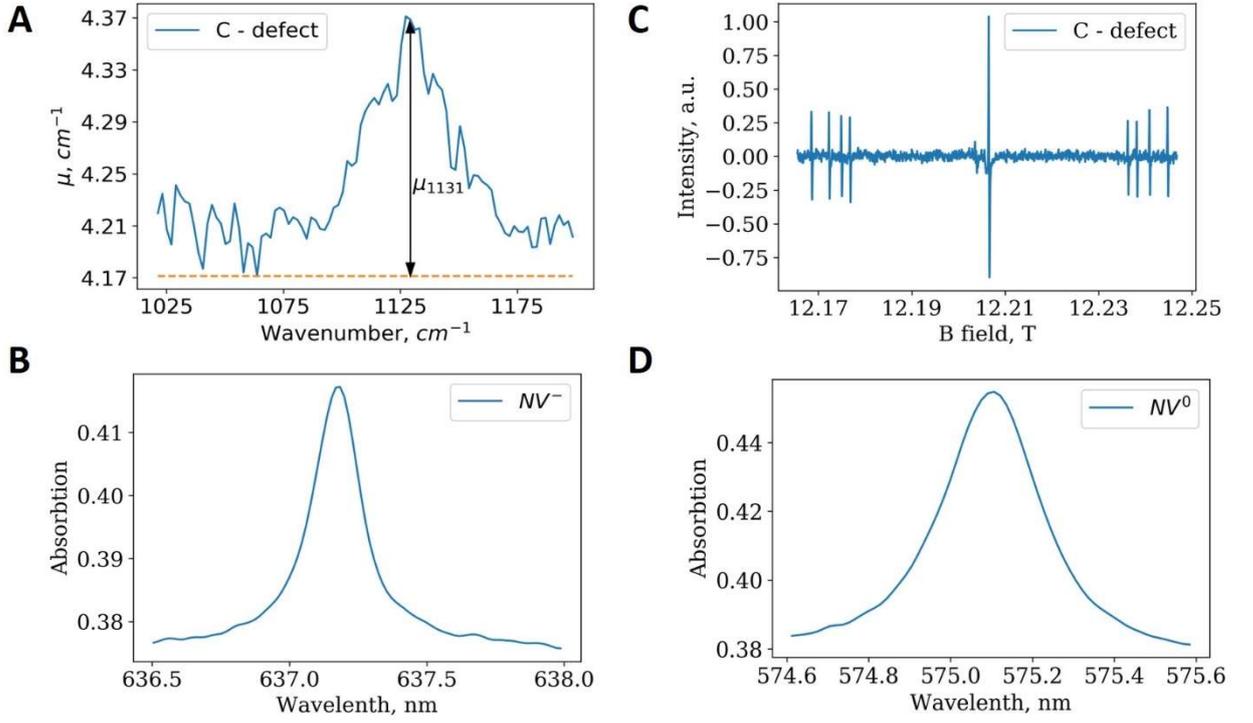

*Figure 1. a) IR absorption spectra of the calibration sample; b) absorption spectra of the $NV^-$ center at cryogenic temperature; c) EPR spectra of the donor nitrogen centers in the calibration sample; d) absorption spectra of the $NV^0$ center at cryogenic temperature; .*

### Coherent property measurements

The dephasing time $T_2^*$ of the $NV^-$ centers was estimated by the fitting decay of the Ramsey fringes[16]. To implement the $\pi/2-\tau-\pi/2$ sequence, we used a combination of microwave and optical pulses that were formed by a radio frequency switch and acousto-optical modulator (see Figure 3B) and have been described in detail elsewhere [6,17,18].

For fitting fringes, we used following equation:

$$F(\tau) = e^{-\left(\frac{\tau}{T_2^*}\right)^2} \sum_{i=1}^{3} C_i \cos(2\pi\Delta_i\tau + \varphi_i) \qquad (7)$$

The coherence time $T_2$ used the same setup with the $\pi/2-\tau-\pi-\tau-\pi/2$ microwave sequence and fitting decay included in the following equation:

$$F(\tau) = Ce^{-\left(\frac{2\tau}{T_2}\right)^\alpha} \qquad (8)$$

Here, we used a high magnetic field of – 80 gauss and determined the revival of the interactions of $^{13}C$ with NV ensembles[19].

# Results

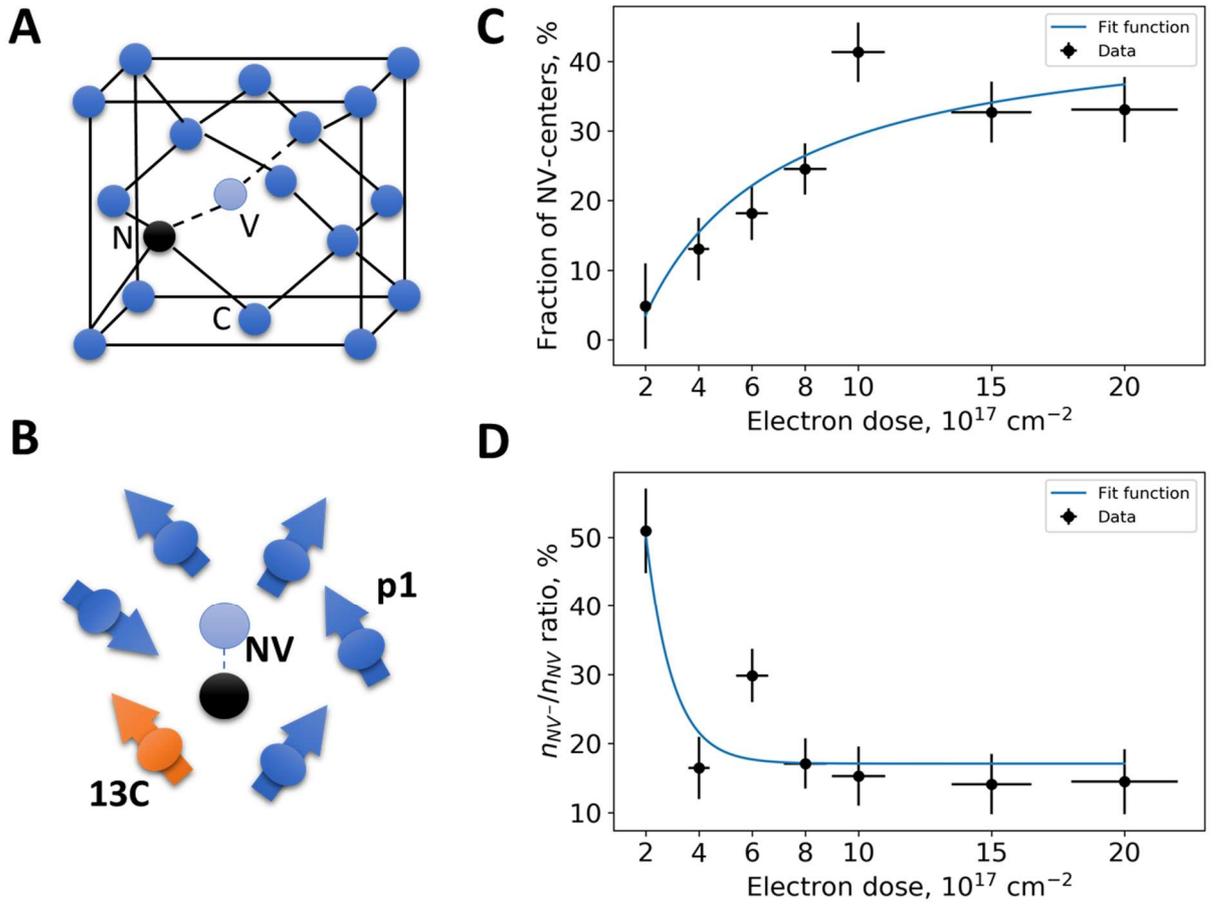

*Figure 2. a) NV center in the crystal diamond structure; b) spin bath of the NV center; c) effect of the electron irradiation doses on the ratio between all the NV centers and all the measured nitrogen defects.; d) effect of the electron irradiation doses on the ratio between the NV centers in the negative charge state and all the NV centers.*

Figure 2C demonstrates how the fraction on NV centers (both, neutral and negatively charged) changes with the electronic irradiation dose of HPHT diamond samples. The dose was varied in a range of $2 \cdot 10^{17}$ to $20 \cdot 10^{17}$ cm$^{-2}$ while maintaining an electron energy of 3 MeV. The saturation behavior of the curve with a fraction of NV centers $\gamma$ can be determined with the following equation:

$$\gamma = \frac{n_{NV}}{n_{NV} + n_C}. \tag{9}$$

The saturation level was found to be 48% using the simple fitting model $f(x)$ as follows:

$$f(x) = \frac{\alpha x}{x + \beta} \tag{10}$$

where $\alpha, \beta$ are fitting parameters, and in our experiments, the best fitting parameters are $\alpha = 57$, $\beta = 10^{18}$ cm$^{-2}$; $n_{NV}$ – is the concentration of all the NV centers; $n_C$ is the concentration of the substitutional nitrogen donor; and $x$ is the electron dose in the irradiation process for each sample.

We note that the fraction ratio (10) does not include other nitrogen related color centers in the diamond (for example C+[15], A[20] and B[8] – defects) concentration, which is difficult to measure at levels of less than $10^{17} \text{cm}^{-3} \approx 1\text{ ppm}$. Therefore, the overall conversion efficiency of nitrogen into NV centers is difficult to observe. This fraction is a convenient parameter, however, and helps to understand the coherence properties of $\text{NV}^-$ centers. NV center fraction saturation occurs at an electron irradiation dose of $15-20 \cdot 10^{17} \text{ cm}^{-2}$ (Figure 2C); therefore, for the nitrogen concentration under consideration, the dose of approximately $15 \cdot 10^{17} \text{ cm}^{-2}$ is optimal.

The dephasing time is expected to be closely related to the donor nitrogen concentration[2]. Indeed, the experimentally measured dephasing time (Figure 3C) decreases with the concentration of the substitutional nitrogen donor $n_C$. The average decrease is approximately $\alpha/x+\beta$, where $\alpha = 0.13$ and $\beta = 0.38$ are fitting parameters. Thus, conversion of a substitutional nitrogen donor into an NV center helps to significantly improve the coherent properties of the $\text{NV}^-$ centers.

Unfortunately, but not surprisingly, the Hann-echo coherence time in not sensitive to the substitutional nitrogen concentration in our range, as shown in Figure 3D.

Figure 2D demonstrates a different parameter – fraction $\kappa$ of useful $\text{NV}^-$ centers $n_{NV^-}$ in all the charged states of the $\text{NV}$ centers, calculated as follows:

$$\kappa = \frac{n_{NV^-}}{n_{NV^-} + n_{NV^0}} . \qquad (11)$$

Here, $n_{NV^0}$ is concentration of the $\text{NV}^0$ color centers. While the coherence time is correlated to a high degree with the presence of neutrally charged NV centers that do not affect the coherence properties, the time does affect the absorption properties of the diamond sample. Much like $\text{NV}^-$, $\text{NV}^0$ absorb excitation light (typically approximately 532 nm) and therefore considerably reduces the utilization of excitation light and the corresponding efficiency of the sensor. By fitting our experimental observations, we found that $\kappa$ could be approximated as $g(x)$ as follows:

$$g(x) = Ce^{-x} + \zeta , \qquad (12)$$

where $C$ and $\zeta$ are fitting parameters, and the best fitting parameters found were $C = 247$ and $\zeta = 17$. At a dose of $15 \cdot 10^{17} \text{ cm}^{-2}$, only 15% of all the NV centers are actually $\text{NV}^-$.

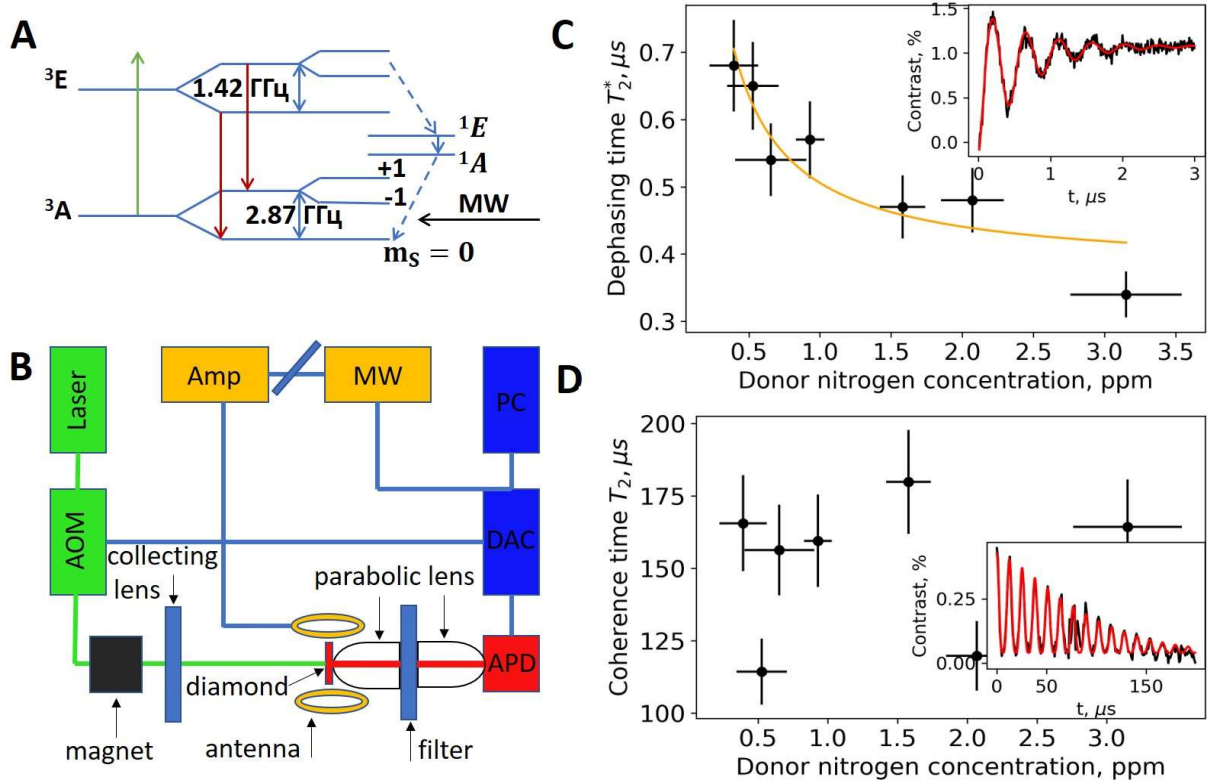

Figure 3. a) Energy system of $NV^-$ centers; b) setup for investigating the NV center spin properties; c) dependence of the dephasing time $T_2^*$ of the NV center ensemble on the substitutional nitrogen donor concentration in the sample; d) dependence of the coherence time $T_2$ of the NV center ensemble on the substitutional nitrogen donor concentration in the sample.

Table 1. All parameters for diamond samples

|  | $T_2$, μs | $T_2^*$, μs | Electron dose, $10^{17}$ cm$^{-2}$ | $NV^-$, ppb | $NV^0$, ppb | $p_1$, ppm |
|---|---|---|---|---|---|---|
| 2V | 164.30 | 0.34 | 2 | 81.8±3.4 | 78.7±2.9 | 3.15±0.39 |
| 4V | 179.88 | 0.47 | 4 | 39.0±2.8 | 198.0±6.0 | 1.58±0.16 |
| 6V | 119.77 | 0.48 | 6 | 137.5±4.9 | 323.3±9.9 | 2.07±0.22 |
| 8V | 159.50 | 0.57 | 8 | 51.8±2.8 | 250.7±6.9 | 0.93±0.10 |
| 10V | 165.57 | 0.68 | 10 | 42.4±2.8 | 235.6±6.9 | 0.40±0.17 |
| 15V | 156.30 | 0.54 | 15 | 44.9±2.8 | 273.1±7.9 | 0.65±0.25 |
| 20V | 114.33 | 0.65 | 20 | 37.7±3.0 | 223.0±7.2 | 0.53±0.18 |

The experimentally measured parameters are summarized in Table 1.

## Discussion

Figure 2 clearly shows that the conversion of donor nitrogen to NV centers strongly depends on the electron dose during the irradiation process in diamond manufacturing. Moreover, the balance between the two different charge states of NV centers sharply changes with increasing electron dose. In this way, one can find the optimal value of the electron dose. The optimum value would, of course, depend on potential applications. For the case of DC magnetometry, sensitivity is proportional to the efficiency of conversion of the excitation light power to the $\text{NV}^-$ center emission and inversely proportional to the square root of the coherence time[1,6]. Thus, in terms of the factors studied in this paper, the sensitivity scales as $\kappa\gamma/\sqrt{T_2^*}$, which allows for an optimal dose of $.10^{18}\,\text{cm}^{-2}$.

At the optimal irradiation dose, the dephasing time is $0.7\ \mu\text{s}$, which is very close to the carbon 13 related decoherence time[3] (approximately $0.9\ \mu\text{s}$). Thus, optimal postprocessing can significantly improve diamond decoherence. In terms of potential sensitivity, this postprocessing procedure gives approximately 2 times better results than nonoptimal processes.

## Conclusions

Intermediate concentration levels (approximately $10^{17}\,\text{cm}^{-3}$) were investigated in terms of the relative concentrations of nitrogen related defects and the coherence properties of $\text{NV}^-$ centers. It was found that the increase of the dephasing rate due to an increasing nitrogen concentration could be compensated via postprocessing procedures. Thus, mid-concentrated nitrogen diamond plates can be efficiently used in sensing applications, in which a long coherence time and relatively large color center concentration is crucial. Compensation results from the conversion of nitrogen donors into neutral NV centers and thus does not solve the problem of sample absorption.

## Acknowledgments

The investigation of NV centers concentrations was performed at the Shared-Use Equipment Center of the TISNCM.

This work was supported in part by the Government of the Russian Federation (Mega-grant # 14.W03.31.0028) for the EPR measurement experiments; in part by the Russian Science Foundation (grant # 18-72-00232) for the nitrogen donors concentration measurements; and in part by the Russian Science Foundation (grant # 19-19-00693) for the coherent diamond property investigation.